\documentstyle[prb,aps,multicol]{revtex}
  \renewcommand{\narrowtext}{\begin{multicols}{2} \global\columnwidth20.5pc}
  \renewcommand{\widetext}{\end{multicols} \global\columnwidth42.5pc}

 \input{psfig}

\renewcommand{\Re}{\mathop {\mathrm {Re}}}
\renewcommand{\Im}{\mathop {\mathrm {Im}}}
\begin{document}
\draft


\title{High-frequency hopping conductivity
in the quantum Hall effect regime: Acoustical studies}

\author{I. L. Drichko$^a$, A. M. Diakonov$^a$, I. Yu. Smirnov$^a$,
Yu. M. Galperin$^{a,b}$, and A. I. Toropov$^c$ }
\address{$^a$A. F. Ioffe Physico-Technical Institute of Russian
Academy of Sciences,
Polytechnicheskaya 26, 194021, St.Petersburg, Russia;\\
$^b$Centre for Advanced Studies, Drammensveien 78, 0271 Oslo, Norway, and
Department of Physics, University of Oslo, P.~O.~Box 1048
Blindern, 0316 Oslo, Norway;\\
$^c$Semiconductors Physics Institute of Siberian Division of
Russian Academy of Sciences, Ak. Lavrentieva 13, 630090,  Novosibirsk, Russia}

\date{\today} \maketitle

\begin{abstract}
The high-frequency conductivity of Si $\delta$-doped GaAs/AlGaAs
heterostructures is studied in the integer quantum Hall effect
(QHE) regime,  using acoustic methods. Both the real and the imaginary
parts of the complex conductivity are determined from the
experimentally observed magnetic field and temperature 
dependences of the velocity and the attenuation of
a surface acoustic wave. It is
demonstrated that in structures with carrier density
$(1.3-2.8)\times 10^{11}$ cm$^{-2}$  and mobility  \mbox{$(1-2)\times
10^5$~cm$^2$/V$\cdot$s} the mechanism of low-temperature
conductance near the QHE plateau centers is hopping. It is also
shown that at magnetic fields corresponding to filling factors
2 and 4, the doped Si $\delta$-layer efficiently shunts the
conductance in the two-dimensional electron gas (2DEG)
 channel. A method to separate the two
contributions to the real part of the conductivity is developed, and
the localization length in the 2DEG channel is estimated.
\end{abstract}
\pacs{PACS numbers: 72.50.+b; 73.40.Kp}
\narrowtext

\section{Introduction}

It is well known from numerous low-temperature resistivity
measurements that the electronic states of a two-dimensional electron
gas (2DEG) whose energies are located between two adjacent Landau
levels in a perpendicular magnetic field are localized.
Consequently, the conductance is determined by electron hopping
between the localized states. The hopping mechanism is 
temperature-dependent. At temperatures 1-4~K, the
conductance $\sigma$ is usually determined by nearest-neighbor
hopping. In this case its temperature dependence is mainly
exponential, $\sigma_{xx}(T)\propto \exp(-E/k T)$ where $E$ is the
temperature-independent activation energy, see e. g.
Refs.~\onlinecite{1,2,3} and references therein.  (We assume
that the 2DEG is located in the $x-y$ plane, the 
magnetic field is parallel to the
$\bf z$-axis, and the electric field is along the $\bf
x$-axis.)
At lower temperatures, $T \lesssim 1$ K, the electron hopping
distance appears to be greater than the typical distance between
nearest neighbors: at such low temperatures it becomes difficult
to find a neighbor whose energy is close to the initial one within
the accuracy $k T$. In this so-called
\emph{variable-range-hopping} regime the conductivity
$\sigma_{xx}$ is also exponentially small,\cite{2,4,5,6} but
with an 
effective activation energy $E$ which is temperature-dependent.

To clarify the nature of the localized states,  
 we study in this paper
the two-dimensional {\em high-frequency} (hf) conductance of a
2DEG,  $\sigma_{xx}(\omega)$, by measuring the
 velocity and the attenuation of surface
acoustic waves
(SAW) propagating along the $x$-direction
nearby the electron layer. Acoustic methods are particularly
promising for our task since $|\sigma_{xx} (\omega)|$ in the
hopping regime may be of
the same order of magnitude as the SAW velocity $V$. Because the
screening of the electric 
fields produced by the SAW is determined by the ratio
$\sigma_{xx}/V$, the acoustic properties
are sensitive to the variations in $\sigma_{xx}$.  Furthermore, 
 the attenuation and the
velocity of the SAW depend on the \emph{complex} conductance,
$$\sigma_{xx} (\omega) \equiv \sigma_1 (\omega)-i \sigma_2 (\omega),$$
and hence both the active, $\sigma_1$, and 
the reactive, $\sigma_2$, components can be
detected. The active component can be then compared to the
static conductivity, $\sigma_{\text{dc}}$. A pronounced
difference will clearly indicate that the electron states are
 localized. We compare the
experimental results for   $
 \sigma_1(\omega)$ and  $\sigma_2(\omega)$  with existing models for 
the dielectric response of localized
 states and extract relevant parameters of the latter.

The paper is organized as follows. In Sec.~\ref{es} the experimental
 setup is presented. Experimental results and their discussion are
 given in Sec.~\ref{er}. They are summarized in Sec.~\ref{cc}. 
Details of the derivation of the expressions 
we use are presented in the Appendix.

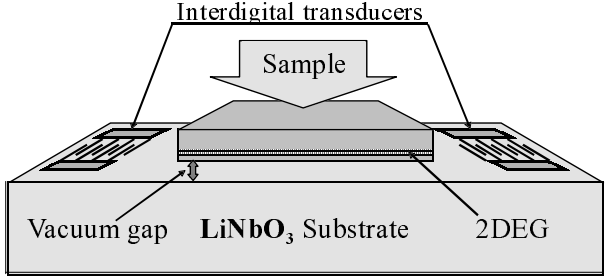
\section{The experimental setup  } \label{es}

A sketch of the experimental setup is shown in Fig.~\ref{f_1}. A SAW
propagating along the surface of a piezoelectric crystal is
accompanied by a wave of hf electric field. This electric field
penetrates into a 2DEG located 
in a semiconductor heterostructure mounted on
the surface. The field produces electrical currents which, in
turn, cause Joule losses. As a result, there are electron-induced
contributions both to the SAW attenuation and to its 
velocity.
These effects are governed by the complex frequency-dependent
conductivity, $\sigma_{xx}(\omega)$, which oscillates as a
function of the external magnetic field. Hence, specific oscillations
will appear both in the SAW attenuation and its velocity. 
Under reasonable
assumptions, the experimental results allow one to determine
$\sigma_{xx}(\omega)$ as  function of the magnetic field and to
analyze its properties.
\input{f_1.inp}
In the present work, the attenuation coefficient $\Gamma$ and the
relative velocity change, $\Delta V/V$, are measured as functions of
perpendicular magnetic field up to 7 T in the temperature interval 1.5-4.2 K. 
The 
Si $\delta$-doped GaAs/AlGaAs heterostructure samples with sheet
densities $n=(1.3-2.7) \times 10^{11}$
 cm$^{-2}$ and mobilities $\mu=(1-2)\times 10^5$ cm$^2$/V$\cdot$s 
at T=4.2 K were grown
by molecular-beam epitaxy, their structures being shown in
Fig.~\ref{structhf}.

\section{Experimental results and their discussion}\label{er}

The expressions relating  $\Gamma$ and $\Delta V/V$ to the complex
conductance,  $\sigma_1-i\, \sigma_2$, can be extracted from
Refs.~\onlinecite{7} and \onlinecite{8}. They read 
\begin{eqnarray}
\Gamma\, \text{(dB/cm)}&=&4.34 \,A K^2 k\,\gamma \, ,\label{Gamma} \\
\Delta V/V &=& (A K^2/2)\,(\delta v/v)\, . \label{Vel}
\end{eqnarray}
Here $k=\omega/V$ is the SAW wave vector, $K^2$  is the
piezoelectric coupling constant of the
substrate (Y-cut LiNbO$_3$), and
\begin{eqnarray}
A&=&8 b(k)(\varepsilon_1+\varepsilon_0) \varepsilon_0^2 \varepsilon_s
e^{-2 k(a+d)} \, , \label{A} \\
b(k)&=&\left( b_1(k)[b_2(k)-b_3(k)]\right)^{-1}\, , \label{b} \\
b_1(k)&=&(\varepsilon_1+\varepsilon_0)(\varepsilon_s+\varepsilon_0)\nonumber \\ &&
-(\varepsilon_1-\varepsilon_0)
(\varepsilon_s-\varepsilon_0)e^{(-2ka)}\, , \label{b1} \\
b_2(k)&=&(\varepsilon_1+\varepsilon_0)(\varepsilon_s+\varepsilon_0)
\nonumber \\ && +
(\varepsilon_1+\varepsilon_0)(\varepsilon_s-\varepsilon_0)e^{(-2kd)} \, ,
\label{b2} \\
b_3(k)&=&(\varepsilon_1-\varepsilon_0)(\varepsilon_s-\varepsilon_0)e^{(-2ka)}
+(\varepsilon_1-\varepsilon_0)
\nonumber  \\ & & \times (\varepsilon_s+\varepsilon_0)e^{[-2k(a+d)]}\,.\label{b3}
\end{eqnarray}
In these equations, $\varepsilon_1$=50, $\varepsilon_0$=1 and
$\varepsilon_s$=12
are the dielectric  constants of LiNbO$_3$, of the vacuum and of the
semiconductor, respectively. $a$ is the finite vacuum clearance
between the sample surface and the LiNbO$_3$ surface, which can be
determined from 
acoustical measurements.\cite{9} $d$ denotes the finite distance 
between the
sample surface and
the 2DEG layer. The parameters of our experimental set are  $a=0.5 \,
\mu\text{m}$,
$d=0.59 \, \mu\text{m}$ for the first sample, 
and $0.09 \mu\text{m}$ for
the second one. The reduced attenuation $\gamma$ and velocity
variation $\Delta V/V$ are given 
by the expressions 
\begin{eqnarray}
&&\gamma =\frac{\Sigma_1}{\Sigma_1^2 +(1+\Sigma_2)^2}\, , \quad
\Delta V/V =\frac{1+\Sigma_2  }{\Sigma_1^2 +(1+\Sigma_2)^2}\, ; \label{atvel}\\
&&\Sigma_{i}= (4\pi \sigma_{i}/\varepsilon_s V)\,t(k), \quad
t(k)=[b_2(k)-b_3(k)]/2b_1(k)\, . \nonumber
\end{eqnarray}
\input{f_2.inp}
The experimental magnetic field dependence of $\gamma$ and  $ \Delta
V/V $ for the
sample with carrier density \mbox{$n=2.7 \times 10^{11}$~cm$^{-2}$}
and mobility $\mu=2\times 10^5$~cm$^2$/V$\cdot$ s are shown in
Fig.~\ref{gvonh}. Similar results have been found previously
in GaAs/AlGaAs heterostructures.\cite{10}

The real and the imaginary parts of the 
complex conductance are derived from $\Gamma$ and $\Delta V/V$,
using Eqs.~(\ref{Gamma})
and (\ref{Vel}). The
results obtained for $T=1.5$ K and acoustic frequency 30 MHz are
shown in Fig.~\ref{s12onh}.
\input{f_3.inp}

\input{f_4.inp}
As can be seen, $\sigma_2$ practically vanishes near half-integer
filling factors, i. e. when the Fermi level is close to any Landau
level. In such regions $\sigma_1 (\omega)$, as has been shown in
 Ref.~\onlinecite{11}, appears to be close to the static 
conductivity,
$\sigma_{\text{dc}}$. These facts indicate that the electron
states are indeed {\em extended}.
 With a further increase of the
magnetic field the Fermi level leaves the Landau band, a
metal-dielectric transition takes place, and the electrons become
localized in the randomly fluctuating potential of the charged
impurities. 

As the Fermi level departs from the Landau level center, $\sigma_1 (\omega)$
becomes clearly larger than $\sigma_{dc}$, see
Fig.~\ref{OldLocal}. Such a behavior can be qualitatively interpreted\cite{11}
as absorption by large clusters (``lakes'') disconnected
from each other. Inside each cluster the absorption is determined by
the value of $\sigma_{dc}$. Since the area occupied by the clusters is
less than the area occupied by the infinite cluster at the mobility
edge, the effective
$\sigma_1 (\omega )$ is less than $\sigma_{dc}$ at half-integer $\nu$.
At the same time, $\sigma_1 (\omega )$ is greater than $\sigma_{dc}$
in the same magnetic field because there is no infinite conducting
cluster at the Fermi level.
The imaginary part, $\sigma_2 (\omega)$ \emph{increases} as the Fermi level
departs from the Landau level's center.   

At magnetic fields corresponding to small integer filling factors,
when the Fermi level finds itself in-between the
adjacent Landau levels, where $\sigma_{dc} \approx 0$, $\sigma_2 (\omega)$
becomes about an order of magnitude larger than $\sigma_1$, see
Fig.~\ref{s12_nu2}.
\input{f_5.inp}
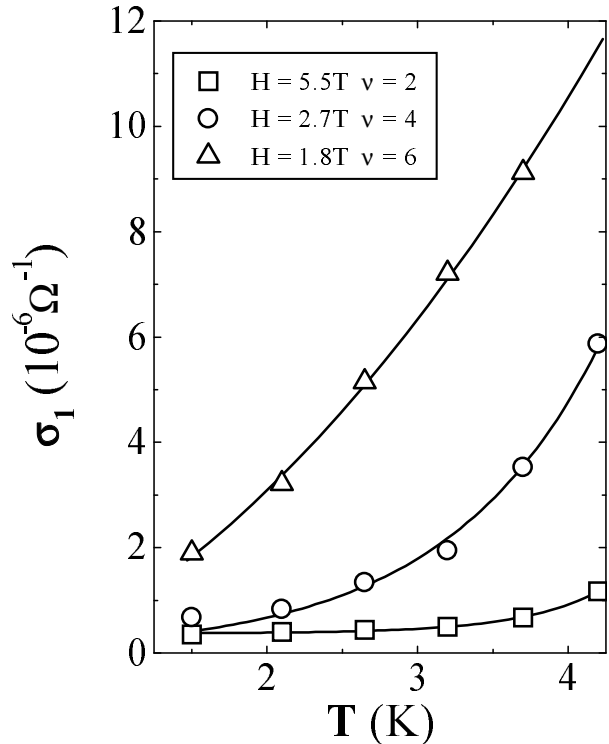
Figure~\ref{s12ont} depicts the temperature dependences of
$\sigma_1(\omega)$ at $f$=30 MHz in magnetic fields corresponding
to the mid-points of the Hall plateaus. One can see a crossover from
a smooth temperature dependence at a strong magnetic field (5.5 T),
to a rather steep increase with  temperature at weaker fields.
Such behavior is compatible with the idea that the conductivity
consists of two contributions. The first one is due to the
extended states near the adjacent upper Landau level, while the
second is coming from the localized states at the Fermi
level.~\cite{12} The relative occupation of the extended states
increases with increasing temperature because of thermally
activated processes. 
Obviously, this effect is dominant at small
magnetic fields.

We now turn to the region of low temperatures and filling
factors close to 2, where hopping between the localized states gives
the main contribution to dielectric response. To analyze the
experimental results we adopt the so-called {\em two-site
approximation}, according to which an electron hops between 
states with close energies localized at two different impurity
centers. These states form {\em pair complexes} which do not
overlap. Therefore, they do not contribute to the static
conductivity but are important for the ac response.
\input{f_6.inp}
Being a very simple, the two-site model has been extensively studied,
see for a review Refs.~\onlinecite{13,14,15} and references therein.
In the following we will use the 2D version of the theory.~\cite{13}
Some details of the discussion depend on the assumptions
regarding both the density of
localized states and the relaxation mechanisms of their population.
We therefore re-derive the theoretical results in Appendix~\ref{ap}.

As is well known \cite{14},  there are two specific
contributions to the high-frequency absorption. The first contribution,
the so-called {\em resonant}, is due to direct
absorption of microwave quanta accompanied by inter-level
transitions. The second one, the so-called {\em relaxational}, or {\em
phonon-assisted}, is due to phonon-assisted transitions which
lead to a lag of the levels populations with respect to the
microwave-induced variation in the inter-level spacing. The
relative importance of the two mechanisms depends on the frequency
$\omega$, the temperature $T$, as well as on sample parameters. The
most important of them is the relaxation rate $\gamma_0 (T)$ of
{\em symmetric} pairs with  inter-level spacing $E=kT$. At
$\omega \lesssim  \sqrt{kT \gamma_0/\hbar}$ the relaxation
contribution to $\sigma_1(\omega)$ dominates, and only this one
will be taken into account. Following the derivation given in 
Appendix~\ref{ap} we obtain
\begin{equation}
  \sigma_1=\frac{\pi^2}{2}\frac{g^2 \xi^3 \omega
  e^4}{\varepsilon_s} ({\cal L}_T +{\cal L}_\omega/2)^2 \, .
  \label{eq3}
\end{equation}
Here  $g$ is the (constant) single-electron density of states at the
Fermi level, $\xi$ is the localization length of the electron
state, ${\cal L}_T =\ln {J/kT}$, $J$ is a typical value of the
energy overlap integral which is of the order of the Bohr energy,
while  ${\cal L}_\omega= \ln(\gamma_0/\omega)$. Eq.~(\ref{eq3}) is
valid provided that the logarithmic factors are large. Note that the
product $r_\omega=\xi({\cal L}_T +{\cal L}_\omega/2)$ is the
distance between the sites forming a hopping pair. Note also that
(\ref{eq3}) is similar to the result obtained in
Ref.~\onlinecite{13}, but differs from it by some logarithmic
factors and a numerical factor of 1/4.

\input{f_7.inp}
The analysis of $\sigma_2 (\omega)$ is a bit more complicated because
virtual zero-phonon transitions give a comparable contribution.
The analysis presented in Appendix~\ref{ap} leads to the following
expression for the ratio $\sigma_2 (\omega)/\sigma_1(\omega)$,
\begin{equation}
\frac{\sigma_2}{\sigma_1}=\frac{2{\cal L}_\omega ( {\cal
L}_T^2 + {\cal L}_T {\cal L}_\omega/2 + {\cal L}_\omega^2/12
)+ 4c {\cal L}_T^2\, {\cal L}_c}{\pi ( {\cal L}_T^2 + {\cal L}_T
{\cal L}_\omega + {\cal L}_\omega^2/4 )}\, . \label{rat1}
\end{equation}
Here ${\cal L}_c=\ln( \hbar \omega_c/kT)$, $\omega_c$ is the
cyclotron frequency, and $c \gtrsim 1$ is a numerical factor
depending on the density of states in the region between the
Landau levels, see Appendix~\ref{ap}. Using the estimate for
$\gamma_0$ from Ref.~\onlinecite{15}, $$\gamma_0=\frac{4 \pi e^2 K^2
kT}{\varepsilon_s \hbar^2 V},$$ valid for the piezoelectric
relaxation mechanism, as well as other parameters relevant to the
present experiment, one concludes that in the hopping regime
$\sigma_2 \gtrsim \sigma_1$. This conclusion agrees with the
experimental results obtained for
 the middles of the Hall plateaus  at 5.5 T and 2.7 T and ensures
that the conductance mechanism in these regions is indeed hopping.

Given an experimental value for $\sigma_1$,  one can obtain from
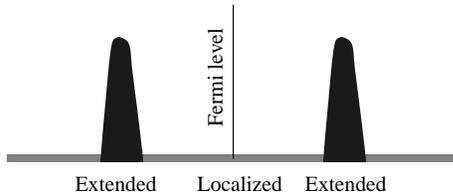
Eq.~(\ref{eq3}) the localization length $\xi$ provided that the
single-electron density of states, $g$, is known for given values
of the magnetic field. This quantity has been obtained from the
temperature dependence measurements of the thermally-activated dc
conductivity.~\cite{1,3}  It has been shown that for small
filling factors the density of states in the plateau regions is
finite and almost field-independent, see Fig.~\ref{saw}
\input{f_8.inp}
Using  the density of states versus mobility curve from
Ref.~\onlinecite{1}, obtained for a sample similar to ours, we
estimate the density of states as $g=2.5\times
10^{24}$~cm$^{-2}\cdot\,$erg$^{-1}$. On the other hand, according to
Ref.~\onlinecite{3},  the density of states as  function of the
magnetic field $H$ can be expressed by the interpolation formula
\begin{equation}
g(H)=\frac{g_0}{1+\sqrt{\mu H}}\, , \label{est3}
\end{equation}
where $\mu$ is the mobility of the 2D-electrons while $g_0=m/(\pi
\hbar^2)$ is the 2D density of states at $H=0$. {}From
Eq.~(\ref{est3}) we obtain for $H=5.5$ T the density of states
$g=1.7 \times 10^{24}$~cm$^{-2}\cdot\,$erg$^{-1}$.

Using the first estimate for the density
of states  one obtains $\xi=6.5 \times 10^{-6}$ cm, that is about
1.6 times greater than the spacer thickness, $l_{\text{sp}}=4
\times 10^{-6} cm$. On the other hand, it is the spacer width
which characterizes the random potential correlation length in the
2DEG layer. Hence, this fact contradicts  our interpretation of
experimental results  in terms of pure nearest-neighbor pair
hopping.

To solve the discrepancy, we assume that the high-frequency hopping
conductivity of the 2DEG channel is shunted by hopping along the
doping  Si $\delta$-layer.
This assumption can be substantiated as follows. Let us suppose
that at the middle of the Hall plateau
$\sigma_1^{\nu=2}=4\times10^{-7}$~Ohm$^{-1}$ and
$\sigma_2^{\nu=2}=2.4 \times10^{-6}$~Ohm$^{-1}$ are entirely
determined by the hopping conductivity along  the  Si
$\delta$-layer. Such a contribution is only weakly dependent on
the magnetic field because the latter is too weak to deform
significantly the wave functions of the Si-dopants. Then the
contributions to $\sigma_i$ associated with the 2DEG channel are
just the difference between the experimentally measured $\sigma_i$
in a given magnetic field  and its value at $\nu=2$.

We now analyze  the dependence of the
 differences $ F_1 \equiv
 \sigma_1 -  \sigma_1^{\nu =2}$ and $F_2 \equiv \sigma_2 - \sigma_2^{\nu
=2}$ on the filling factor $\nu$. The plots 
of $\lg F_i$ versus $\nu$
are shown in Fig.~\ref{ksionnu}. Both curves approach straight lines, 
and consequently can be extrapolated to $\nu =2$.
Using this extrapolation we have obtained $F_1^{\nu
=2}=10^{-8}$~Ohm$^{-1}$ and $F_2^{\nu =2}=5 \times
10^{-8}$~Ohm$^{-1}$.
 It should be noticed here that the extrapolated values of $F_i^{\nu
 =2}$ are two orders of magnitude smaller than the values of
 $\sigma_i^{\nu =2}$, associated with the hopping along
 Si-$\delta$-layer.
\input{f_9.inp}
Using the extrapolated values of $F_1$ and $F_2$ to extract the
2DEG contributions to $\sigma_1$ and $\sigma_2$, one can calculate
the electron localization length at $\nu = 2$ from
Eq.~(\ref{eq3}).
This procedure is corroborated by the fact  that the experimental
ratio $F_2/F_1 = 5$
is close to the theoretical value 4.2 coming from
Eq.~(\ref{rat1}).
 The localization length at $\nu=2$ obtained in this way is $\xi =2
\times 10^{-6}$ cm, which is  half of the spacer width.  This
estimate makes realistic the "two-site model" which we have
extensively used. It should be emphasized, however, that from the
above value of $\xi$ the hopping length $r_\omega$ is estimated to be
$1.4\times 10^{-5}$ cm. Consequently, there is an interplay
between hops to the nearest and more remote neighbors. A more
rigorous theory for this situation should be worked out. Such a
theory should also explain why the magnetic field dependences of
$\sigma_1$ and $\sigma_2$  at the 
vicinity of $\nu$ = 2 appear to be different
-- the $\sigma_1(H)$-dependence is more pronounced than the
$\sigma_2(H)$-one. According to the two-site model, both
are  determined by 
the respective dependence of the localization length on the 
magnetic field and should be
similar. Indeed, their ratio, from Eq.~(\ref{rat1}), is almost
field-independent. It follows from the experimental data that
there exists an additional mechanism leading to the pronounced
decrease of  $\sigma_2$  as the Fermi level falls into the
extended  states region. A probable mechanism is thermal
activation of electrons from the Fermi level to the upper Landau
band, leading, firstly, to a decrease of the number  of pairs
responsible for the hopping conductivity, and, secondly to a
screening of the electric field amplitude produced by the SAW. We
hope to work out a proper quantitative theory in future.

\section{Conclusions}\label{cc}

The above analysis leads to the following conclusions:
\begin{itemize}
\item At the vicinity of the Hall plateau centers, high-frequency
hopping conductance
in the 2DEG layer can be effectively shunted by hopping inside the
doping $\delta$-layer.
\item When the shunting effect is properly subtracted, the results
appear to be compatible with the nearest neighbor
two-site model of hopping
conductivity.
\item The localization length determined 
at different magnetic fields
(and, consequently, different 
filling factors) by the above method scales as the
magnetic length $a_H=(\hbar/eH)^{1/2}$. This agrees with the concept 
of nearest-neighbor hopping.
\item The interpretation of the 
imaginary part of the conductivity, $\sigma_2
(\omega)$, appears more complicated.
 While the magnetic field dependence of the real part of the 
hf hopping conductivity
 of 2D electrons seems to be determined by the
magnetic field dependence of the localization length -- the slope
of $\lg \xi (H)$, calculated from the values
of $F_1^{\nu=2} (H)$ using Eq.~(\ref{eq3}), is close to the slope
of $\lg \xi (H)$ in Ref. \onlinecite{2} --  
the magnetic field dependence of the imaginary part of the
$hf$ conductivity has been explained so far 
only qualitatively.
A more detailed quantitative analysis, which
would include a proper account of the screening of the SAW-induced
hf electrical field by both layers, is required.
\end{itemize}
It is worth emphasizing that the acoustic method used in
the present work allows the determination of the localization length
near the Hall plateau centers.This is very difficult to achieve
using a dc technique.

\section{Acknowledgments}

The work is supported by RFFI N 98-02-18280, MNTRF N 97-1043
grants and I.~L.~Drichko was supported by the grant of Research
Council of Norway. We are grateful to Ora Entin-Wohlman for reading
the manuscript and useful remarks.

 \appendix
\section{Derivation of $\sigma (\omega)$} \label{ap}
The derivations of the complex $\sigma (\omega)$ within the
two-site approximation  have been extensively discussed, see e. g.
Refs.~\onlinecite{14} and \onlinecite{15}. However, the resulting 
formulae differ in some
details. These differences are mainly due to different assumptions
about the relaxation of the occupation numbers of the localized states.
We therefore present here a unified derivation, in order 
to clarify the various
assumptions and notations.

Let us characterize the sites by the single-electron energies
$\varphi_{1,2}$ which would be the actual energies if the Coulomb
correlation between the occupation numbers is ignored. The two
electron energies, in the absence of quantum hybridization of the
states, can be specified by 4 terms, $$W_0=0,  \quad W_1=\varphi_1,
\quad W_2=\varphi_2, \quad W_3=\varphi_1+\varphi_2
+\frac{e^2}{\varepsilon_s r}\, .$$ Here $r$ is the distance
between the sites. As shown in Ref. \onlinecite{14}, 
when the frequency $\omega$
and the temperature $T$ are low enough, such that
$\hbar \omega, \ kT \ll e^2/\varepsilon_s r$
only the two terms $W_1$ and $W_2$ can be occupied, and we face a
situation of a two-level electronic system (TLS). Quantum tunneling
hybridizes the two levels, so that the resulting energies are
$$W_\pm=\frac{\varphi_1 +\varphi_2}{2}\pm \frac{E}{2}\, , \quad E=
 \sqrt{\Delta^2 + \Lambda^2 (r)}\, .$$
Here $\Delta=\varphi_1-\varphi_2$,  $\Lambda (r)$ is the 
energy overlap
integral which decays as $r$
increases. The effective Hamiltonian corresponding to this situation is
\begin{equation}\label{eg00}
{\cal H}_{LR} =\frac{1}{2}\left(\begin{array}{cc} \Delta& - \Lambda\\
-  \Lambda& -\Delta \end{array}\right)=\frac{1}{2} \left(\Delta \, \sigma_z
-\Lambda\,  \sigma_x \right)\, .
\end{equation}
Here $\sigma_i$ are the Pauli matrices.
Diagonalizing this Hamiltonian we obtain
${\cal H}_0 = (E/2)\, \sigma_z$.

The quantities $\varphi_i$ are random, and their distributions are
specified as follows.\cite{14} The ``center-of-gravity'',
$(\varphi_1+\varphi_2)/2$ is assumed to be uniformly distributed 
within a band
of width $e^2/\varepsilon_s r$; and the
difference, $\Delta \equiv \varphi_1-\varphi_2$, is also assumed to be
uniformly distributed within a band much wider than $kT$. 
Since $d^2r =r\, dr\, d\phi$ where
$\phi$ is the polar angle in the 2DEG plane, we obtain an
$r$-independent pair distribution function
${\cal P} (\Delta,r,\phi)=g^2e^2/\varepsilon_s$
where $g$ is the (constant) single-electron density of states.
 It is convenient to change the variables 
from $\Delta,r,\phi$ to $ \Delta,
\Lambda, \phi$,
\begin{equation} \label{df2}
{\cal P} (\Delta, \Lambda, \phi)=g^2 (e^2/\varepsilon_s)\,
|d r_\Lambda/d
\Lambda|\, ,
\end{equation}
where $r_\Lambda$ is the solution of the equation 
$\Lambda(r)=\Lambda$.

In the presence of an external ac electric field ${\bf E}$ the pair
 acquires a dipole moment ${\hat {\bf d}}=
 e {\hat {\bf r}}$,  described by 
the interaction Hamiltonian  ${\cal
 H}_i=({\bf E}\cdot {\hat {\bf d}})=e({\bf E}\cdot {\hat {\bf
 r}})$.
This interaction is added to  $\Delta$ in the
 Hamiltonian (\ref{eg00}). In the representation where ${\cal
 H}_0$ is diagonal, the interaction Hamiltonian becomes
\begin{equation} \label{hi1}
  {\cal H}_{\text{int}}=e({\bf E}\cdot{\hat {\bf
 r}})\left(\frac{\Delta}{E}\, \sigma_z -\frac{\Lambda
}{E}\, \sigma_x \right) \, .
\end{equation}

The contribution of a pair to the complex $\sigma( \omega)$ can be
expressed in terms of the complex susceptibility, 
$\chi (\omega) =\sigma(
\omega)/i \omega$
which in turn is given by (cf. Ref. \onlinecite{16})
\begin{eqnarray}
\chi(\omega)&=&\frac{\pi e^4\, g^2}{\varepsilon_s}\int d \Delta\, d
\Lambda\, r_\Lambda^2 \, \left|d r_\Lambda/d \Lambda\right|\nonumber \\
&& \times
\left[\left(\Delta/E \right)^2 \chi_{zz} (\omega) +
\left(\Lambda/E \right)^2 \chi_{xx} (\omega)  \right] \, .
\end{eqnarray}
The partial susceptibilities $\chi_{pq}$ are given by 
(cf. Ref. \onlinecite{16})
\begin{eqnarray}
\chi_{zz}&=&\frac{1}{kT\cosh^2(E/2kT)}\, \frac{ i
\gamma_\parallel}{\omega + i \gamma_\parallel} \\
\chi_{xx}&=&\tanh \left(\frac{E}{2 kT} \right)\, \sum_\pm
\mp \frac{ \hbar^{-1}}{\omega \mp E/\hbar + i \gamma} \, ,
\end{eqnarray}
where $\gamma$ and $\gamma_\parallel$ are the proper relaxation rates.
$\chi_{zz}$ is responsible for the relaxational contribution, while
$\chi_{xx}$ is responsible for the resonant one. 

To continue the calculations one needs to 
specify the spatial dependence
of the overlap integral. Let us assume that
$\Lambda(r)=J \, e^{-r/\xi}$,
where $\xi$ is the localization length. Then, $r_\Lambda=\xi\,
\ln (J/\Lambda)\, , \quad |d r_\Lambda/d \Lambda|= \xi/\Lambda$.
 At the
next step, it is convenient to transform the variables from
$\Delta ,\Lambda$ to $E, p=(\Lambda/E)^2$, the Jacobian being
$(2p)^{-1}(1-p)^{-1/2}$. This results in
\begin{eqnarray}\label{res2}
\tilde{\chi}(\omega) \equiv \frac{\chi(\omega)}{\chi_0}&=&\int_0^\infty \! d
 \epsilon \int_0^1  \frac{dp}{p\sqrt{1-p}}  \, \ln^2
 \left(\frac{\tilde J}{\epsilon  \sqrt{p}}\right) \nonumber \\ &&\times
\left[p \,\chi_{xx} (\omega) +
(1-p) \, \chi_{zz} (\omega)  \right] \, 
\end{eqnarray}
where
$\chi_0=\pi e^4\, g^2\, \xi^3 kT/2\varepsilon_s$, $\epsilon = E/kT$,
and ${\tilde J} = J/kT \gg 1$.

The following analysis will be based on Eq.~(\ref{res2}). It can be
easily shown that under the conditions of the present experiment
 the only important contribution to the dissipative part of the
susceptibility, $\Im {\tilde \chi} (\omega)$, is
the one coming from  the relaxational mechanism, $\chi_{zz}$. Thus we have,
\begin{eqnarray} \label{im0}
\Im  {\tilde \chi}(\omega)&=&\int_0^\infty \! \frac{d
\epsilon}{\cosh^2 (\epsilon/2)}  \int_0^1 \! \frac{dp \,\sqrt{1-p}}{p}
\nonumber \\ &&\times
\ln^2 \left(\frac{\tilde J}{\epsilon  \sqrt{p}} \right) \,
\frac{\gamma_\parallel \omega}{\omega^2 + \gamma^2_\parallel}\, .
\end{eqnarray}
An important feature of the  relaxation rate $\gamma_\parallel$
is that $\gamma_\parallel(\epsilon , p) = p\, \gamma_0
(\epsilon)$, see e. g. Ref.~\onlinecite{15}. Since we are
interested in the case $\omega \ll \gamma_0$, we can put
$\epsilon=1, \quad p=p_\omega \equiv \sqrt{\gamma_0(1)/\omega}$ in
the argument of the logarithm and take the logarithm out of
the integrand. As a result,
\begin{equation} \label{im}
\Im {\tilde \chi}=\pi \left({\cal L}_T +\frac{1}{2}{\cal L}_\omega\right)^2 \,,
\end{equation}
where ${\cal L}_T =\ln {\tilde J} \gg1 $ while
${\cal L}_\omega= \ln(1/p_\omega) \gg 1$.

The relaxational contribution to the real part can be written as
\begin{eqnarray} \label{Rez}
&&\Re \, {\tilde \chi}_{zz} (\omega)=\int_0^\infty \! \frac{d \epsilon}{\cosh^2 (\epsilon/2)}
\int_0^1 \frac{dp \,\sqrt{1-p}}{p}
 \ln^2 \left(\frac{\tilde J}{\epsilon  \sqrt{p}} \right)
\nonumber \\ && \quad \times
\frac{\gamma^2_\parallel }{\omega^2 + \gamma^2_\parallel}
\approx 2 \int_{p_\omega}^1 \frac{dp}{p} \left( {\cal L}_T -
\frac{1}{2}\ln p \right)^2 \nonumber \\
&& \quad =2{\cal L}_\omega \, \left( {\cal
L}_T^2 + \frac{1}{2}{\cal L}_T \, {\cal L}_\omega + \frac{1}{12}{\cal
L}_\omega^2 \right)\, .
\end{eqnarray}
The contribution from $\chi_{xx}$ to the real part of the
susceptibility is also important. Putting
$\hbar \omega \ll E$ we obtain
\begin{eqnarray} 
\Re  {\tilde \chi_{xx}}&= &2\int_0^\infty \frac{d \epsilon}
{\epsilon} \tanh (\epsilon/2) \int_0^1  \frac{dp}{\sqrt{1-p}}  \, \ln^2
\left(\frac{\tilde J}{\epsilon \sqrt{p}} \right) \nonumber \\
&\approx& 4 {\cal L}_T^2 \, \int_0^\infty \frac{d \epsilon}
{\epsilon} \tanh \left(\frac{\epsilon}{2}\right) \, .
\end{eqnarray}
The last integral diverges logarithmically at its upper limit. That
means that the result is substantially
dependent on the total structure of the impurity band. Assuming that 
(i) $e^2/\varepsilon_s
\xi {\cal L}_T \ll \hbar \omega_c$, and
(ii) that
we are interested in the situation when the Fermi level is in the
middle of the gap,  we can
replace for a {\em very crude} estimate
$$g^2 \int_0^\infty \frac{d \epsilon}
{\epsilon} \tanh \left(\frac{\epsilon}{2}\right) \quad \text{by} \quad  \int_0^\infty
\frac{d \epsilon}
{\epsilon}\, g^2 (\epsilon) \tanh \left(\frac{\epsilon}{2}\right)\, .$$
According to certain experimental evidence, the density of localized
states within the gap is a weak function of the energy.
Then an
estimate for the above integral can be written as $c\, g^2 {\cal
L}_c$ where
${\cal L}_c=\ln( \hbar \omega_c/kT)$, while $c \gtrsim 1$ is a
correction factor due to energy dependence of the density of states.
As
a result, we obtain
\begin{equation} \label{Rex}
\Re {\tilde \chi}_{xx} \approx 4 c\, {\cal L}_T^2\, {\cal L}_c\, .
\end{equation}
Since $\Re\, \sigma/\Im \, \sigma=\Im \, \chi /\Re \, \chi$ we obtain
Eq.~(\ref{rat1}).

\widetext
 \end{document}